\documentstyle[preprint,aps]{revtex}
\begin{document}
\draft
\title{
Realization of a Spin Liquid in a\\
Two Dimensional Quantum Antiferromagnet}

\author{A. Parola}
\address{Dipartimento di Fisica, Universit\'a di Catania, Catania, Italy.}

\author{S. Sorella and Q.F.Zhong }
\address{International School for Advanced Studies,
Via Beirut 4,34014 Trieste, Italy}
\date{preprint: SISSA 128/93/CM/MB BABBAGE: cond-mat/9308002}

\maketitle
\begin{abstract}
The ground state properties of the
two dimensional spatially anisotropic Heisenberg model are
investigated by use of field theory mappings, spin-wave expansion
and Lanczos technique.
Evidence for a disorder transition induced by anisotropy at about
$J_y/J_x < 0.1$ is shown. We argue that the disordered phase is
gapless and its long wavelength properties
can be interpreted in terms of decoupled one dimensional chains.
\end{abstract}

\narrowtext
The search for disordered (spin liquid) ground states in two dimensional (2D)
electronic models
has been pursued since the seminal work of Fazekas and Anderson\cite{faz}
on quantum antiferromagnets (QAF) in frustrating lattices.
This problem has been revived in the last few years due to the resonating
valence bond conjecture\cite{pwa}
which stimulated much numerical work on the subject. Despite the considerable
effort, the existing evidence in
favor of a spin liquid in 2D frustrated QAF is weak at best\cite{fr,schulz}
with the single exception of the Kagom\'e QAF where a disordered ground state
is plausible\cite{huse},
even if the presence of non conventional magnetic order is still a possibility.

In this Letter, we present analytical as well as numerical evidence
supporting an order--disorder transition in the square lattice $S=1/2$ QAF
driven by spatial anisotropy in the nearest neighbor coupling.
This model does {\it not} introduce frustration and therefore presents
several advantages with respect
to the previously investigated systems, the most important being
the absence of any plausible order parameter competing with the N\'eel
staggered magnetization ${\bf m}=\sum_{\bf R} {\bf S_ R}\,\exp({\bf Q\cdot R})$
$\left [{\bf Q}=(\pi,\pi)\right ]$. The model is defined by the hamiltonian
\begin{equation}
H=\sum_{\bf R}\left[
J_x {\bf S_R\cdot S_{R+x}} +
J_y {\bf S_R\cdot S_{R+y}} \right ]
\label{H}
\end{equation}
where ${\bf S_R}$ are spin $1/2$ operators living on a square lattice,
${\bf x}$ and ${\bf y}$ are unit vectors and $0\le J_y\le J_x$.
The isotropic limit ($J_y=J_x$) has been extensively studied by
exact diagonalizations\cite{oitmaa}
and quantum Monte Carlo\cite{ceperley,runge}
with the resulting evidence of a finite staggered magnetization in
the thermodynamic limit\cite{runge}
$m\sim 0.3075$ quite close to the spin-wave theory (SWT)
estimate $m=0.3034$\cite{oguchi}.
Physically, the strongly anisotropic model (\ref{H}) describes a system of
weakly coupled AF chains whose study has attracted considerable interest
among theoreticians and experimentalists in view of the possibility
to observe the peculiar features of one dimensional physics\cite{bourb}.

The presence of an order--disorder transition in model (\ref{H}) has been
conjectured by several authors\cite{iome,doucot}
and can be motivated by the standard mapping of the 2D quantum model
(\ref{H}) into the (2+1) dimensional $O(3)$ non linear sigma model
(NL$\sigma$M)
defined by the action
\begin{equation}
S={1\over 2 }\int dx dy dt \left[
\Upsilon_x (\partial_x {\bf n})^2 +
\Upsilon_y (\partial_y {\bf n})^2 +
\chi_0 (\partial_t {\bf n})^2 \right ]
\label{S}
\end{equation}
where ${\bf n}$ is a unit vector. The lowest order estimates of the
parameters give $\Upsilon_x=J_x/4$, $\Upsilon_y=J_y/4$,
$\chi_0^{-1}=4a^2(J_x+J_y)$ where $a$ is the lattice spacing.
Two limits of the action (\ref{S}) can be easily analyzed:
The isotropic model is known to be ordered for
the physically relevant parameters\cite{nelson}
while the $J_y\to 0$ limit of the
action (\ref{S}) correctly describes a stack of uncoupled (1+1) dimensional
models which are disordered at any finite ``temperature"
$g=(\Upsilon_x\chi_0 a^2)^{-1/2}$ owing to the Mermin Wagner theorem.
Most interestingly, the order--disorder transition occurs at
a {\it finite} value of the spatial anisotropy and belongs to the
universality class of the classical three dimensional Heisenberg model.

Similar conclusions can be drawn directly
from  SWT  on the quantum hamiltonian (\ref{H}).
In fact, the $1/S$ expansion of the staggered magnetization
can be straightforwardly generalized to anisotropic models and
predicts a breakdown of N\'eel order at about
$\alpha_c\sim 0.03$ ($0.07$) at first (second) order in $1/2S$.
The {\it increase}
of the critical anisotropy parameter $\alpha_c$ in going from first to
second order gives confidence about the actual occurrence of the transition,
which is in fact enhanced by quantum fluctuations.
Therefore, on the basis of the field theory mapping and SWT,
we expect that
by lowering the anisotropy parameter $\alpha=J_y/J_x$ a disordered
phase sets in within a finite interval $\alpha_c >\alpha > 0$. This prediction
should be qualitatively correct because field theory methods are known
to reproduce the physics of QAF both in the isotropic two
dimensional limit\cite{nelson}
and in the one dimensional ($\alpha=0$) case, provided the
topological term is included in (\ref{S})\cite{haldane}.

In order to test the theoretical predictions on the model (\ref{H})
and to determine the properties of the two dimensional
spin liquid state, we have performed Lanczos diagonalizations on
several lattices up to 32 sites. A proper finite size scaling of
small lattice results
is obviously important in order to provide a correct interpretation of
the diagonalization data.
Fortunately, much work has been recently devoted to this
subject\cite{fisher,ziman}
showing that, in the ordered phase,
the renormalization group flow drives the model towards weak coupling
making SWT asymptotically exact at long wavelengths.
As a consequence, SWT  is able to describe the leading
size corrections in finite lattices and therefore represents a
powerful method for analyzing small size data, particularly in
non frustrated models.
SWT, generalized to finite systems\cite{zhong},
compares quite favorably to numerical
results in bipartite lattices. The good agreement also persists for
the anisotropic model (\ref{H}), as shown in Fig. 1 where a finite
size estimate of the order parameter is plotted as a function of
the anisotropy $\alpha$. The breakdown of SWT
at $\alpha < 0.1$ again suggests that a qualitative change in the
ground state properties is occurring in this regime.

Further numerical evidence of the phase transition can be obtained
by the structure  of the energy spectrum as a function of the
total uniform magnetization. According to a recent analysis\cite{fr}
the presence of N\'eel long range order in the thermodynamic limit
reflects in the structure of the energy spectrum in
finite size systems. In fact, if long range antiferromagnetic order
is present in the system, the dependence of the
energy $E(S)$ on the total spin $S$ must follow the approximate relation:
\begin{equation}
E(S)\sim E(0)+ S(S+1)/(2 I_N)
\label{es}
\end{equation}
up to a maximum value
$S_{max}$ of the order of the square root of the number of sites $N$.
Eq. (\ref{es})  approximately reproduces the energy spectrum of a spin-$S$
rigid rotator with a momentum of inertia per site $I_N/N$ corresponding to
the uniform susceptibility $\chi$ of the model.
Notice that this criterion correctly reproduces the absence of
antiferromagnetic long range order in one dimension where
the energy spectrum scales as\cite{haldane}
$E(S)\sim E(0)+ S^2/(2\chi N)$.
The relation (\ref{es}) is in fact quite well verified
both for the isotropic system and for the anisotropic ones up to
$\alpha\sim 0.1$ where significant discrepancies appear (see Fig. 2a).
 The strong deviations from Eq. (\ref{es}) which develop in the numerical
data can be directly related to the asymptotic decoupling of the chains
in the square lattice, leading to an approximate {\it linear} dependence
$E(S)\sim E(0)+\Delta S$.
This anomaly does not seem to scale to
zero in the thermodynamic limit but instead persists in all the lattices
we have analyzed. From our finite size data an accurate estimate of
the uniform susceptibility can be extracted by a quadratic fit of the energy
spectrum $E(S)$ at small but finite (uniform) magnetization $S/N$.
The results can then be extrapolated to infinite volume
by use of a finite size scaling of the form\cite{runge}
$\chi^{-1}=\chi^{-1}_{\infty}+A\,N^{-1/2}+B/N$
which is known to work both in the isotropic limit and in one dimension
(extreme anisotropy). The extrapolation, shown in Fig. 2b for several
values of the anisotropy parameter, is in fact in good agreement with
the known values at $\alpha=0$ and  $\alpha=1$ (also shown in the figure)
and predicts a smooth, featureless
susceptibility in the whole anisotropy range. This result indicates that
the system remains gapless across the phase transition and suggests that
the nature of the disordered phase of this model might be more exotic
than the expected nondegenerate singlet, in agreement with a conjecture
put forward by Haldane\cite{haldane}.

However, the possibility of a gapless phase
contrasts with the commonly accepted
phase diagram of the model (\ref{H}) defined on two chains\cite{strong,dag2c}
where a gap $\Delta(\alpha)$ is believed to open at every finite
value of the anisotropy parameter $\alpha$. The disordered phase
in the two chain model is in fact continuously connected with the
$\alpha\to\infty$ limit where the gap is interpreted as the effect
of the finite size of the lattice along the $y$ direction.
The same result actually holds for every {\it even} number of chains
while an {\it odd} chain model remains gapless all the way to the
$\alpha \to\infty$ limit. Therefore it is not too surprising that
the hamiltonian (\ref{H}) on square clusters preserves the
peculiarities of the odd chain sequence and does not open a gap
at any $\alpha$. In finite clusters, however, a gap is always
present and we must investigate whether it disappears in the
thermodynamic limit. We have analyzed the finite size scaling
of the gap in the case of two chains ($L\times 2$),
three chains ($L\times 3$) with antiperiodic boundary conditions along the
$y$ direction, and for square clusters.
For any $\alpha$ the lowest excited state is always a
triplet but its size dependence
is quite different in the three cases. In order to see whether
a gap is present in the strong anisotropic region we assumed that,
for $\alpha\to 0$ and $L\to\infty$, the gap $\Delta(L,\alpha)$ can be
expressed in a scaling form, as usual near a critical point:
\begin{equation}
\Delta(L,\alpha)=\Delta(L,0)\,F\left [\alpha L\,(\log\,L)^{1/2}\right]
\label{scala}
\end{equation}
where the one dimensional gap $\Delta(L,0)$ is known to scale as
$1/L$. The specific form (\ref{scala}) has been chosen in order to match
with first order perturbation theory in $\alpha$ and does not
depend on the number of chains of our lattice. However, the scaling
function $F(x)$ behaves quite differently in the three geometries,
as can be seen in Fig.3. The correctness of our scaling form (\ref{scala})
can be inferred by the collapse of the finite size numerical data
on a smooth curve in all cases, provided $\alpha$ is sufficiently small.
The region where the universal curve $F(x)$
is defined increases with growing size and the thermodynamic limit at
fixed (small) $\alpha$ corresponds to the large $x$ region of the
scaling curve which should be extrapolated from the finite size data.
In the two chain model $F(x)$ clearly goes through a
minimum and then grows, suggesting a linear asymptotic behavior
at large $x$ which implies a finite gap of order $J_y$ at small $\alpha$
in agreement with field theoretical analysis\cite{strong}.
Instead, the scaling function is always monotonic both in the three
chain case and, even more convincingly,
in the square clusters, supporting the absence of a gap in these systems.

In order to understand how a disordered gapless phase may appear in 2D
it is useful to consider
other physical quantities like the spin-wave velocity
and the momentum dependence of the magnetic structure factor.
Again, SWT  provides a valuable help in the interpretation
of the numerical results. The spin velocity is almost constant
at all anisotropies ranging between the one dimensional value
$c_x=\pi/2$ and the isotropic limit\cite{runge} $c_x\sim 1.56$
which are both reproduced within 10\% by second order SWT
generalized to anisotropic models. A surprising result of
SWT is the enhancement of the anisotropy in the spin
velocity ratio induced by quantum fluctuations:
\begin{eqnarray}
\left({c_y\over c_x}\right )^2&=&\alpha Z(\alpha) \,\,\,\,\,\,
Z(\alpha)=1-{1\over 2S}C(\alpha) \\ \nonumber
C(\alpha)&=&\frac{1}{N}\displaystyle\sum_k~^\prime\frac{(\cos k_x-\cos k_y)
(\cos k_x+\alpha\cos k_y)}{\sqrt{(1+\alpha)^2-(\cos k_x+\alpha\cos k_y)^2}}
\\ \nonumber
\end{eqnarray}
In fact, while at lowest order the spin velocity ratio coincides with
the anisotropy parameter, the one loop calculation always reduces the
$Z(\alpha)$ factor. Obviously, the correction $C(\alpha)$
vanishes at the isotropic
point $\alpha=1$ but diverges logarithmically in the $\alpha\to 0$ limit.
Therefore, SWT  suggests the occurrence of a decoupling transition
at a finite value of $\alpha$ signaled by $Z(\alpha_c)=0$.
The same anisotropy renormalization factor $Z(\alpha)$ governs the
long wavelengths behavior of the physical correlation functions. In
particular, the magnetic structure factor behaves as
\begin{equation}
S(k_x,k_y)\propto \sqrt{k_x^2+\alpha Z(\alpha) k_y^2}
\label{sk}
\end{equation}
In order to verify these
predictions we tested Eq.(\ref{sk}) against Lanczos diagonalizations
in the 32 sites lattice. The results are shown in
Fig. 4 together with the zero and one loop SWT results
for the spin velocity ratio in the thermodynamic limit.
The numerical data are in good agreement with the spin-wave results
in the 32 site lattice, and show an even larger effect.
Therefore we are led to conclude that at long wavelengths a decoupling
transition may actually occur in strongly anisotropic spin models.
The phase diagram of the anisotropic model (\ref{H}) suggested by
SWT is depicted in Fig. 5 for generic spin $S$ systems.
The transition line where the staggered
magnetization vanishes has been calculated
at the lowest order spin-wave level together with
the locus $Z(\alpha)=0$ where we expect a ``decoupling
transition''. At the same order in $1/S$ we have found that these
two lines approximately coincide up to a critical value of $\alpha$
beyond which
the system disorders without decoupling. In the strong anisotropy limit the
transition is characterized by the vanishing of both
the staggered magnetization and the spin velocity ratio leading to a
picture of basically uncoupled chains with interesting experimental
consequences about the possibility to observe 1D behavior in real systems.
We believe that this  phase diagram  is qualitatively correct although
higher order terms in the
SWT  expansion (available only for the magnetization)
may quantitatively change the phase transition line.
In order to fully characterize the disordered phase, topological defects
must be taken into account leading to a possible
difference between integer and half integer spin systems\cite{haldane}.

We gratefully acknowledge E. Tosatti  for
useful discussions, constant encouragement and (AP) warm hospitality
at ISAS.
This work has been partially supported by CNR under Progetto
Finalizzato ``Sistemi informatici e calcolo parallelo''.

\begin{figure}
\caption{ Order parameter
$m=\protect\sqrt{S(\pi,\pi)/V }$ for different
lattice sites compared with the first (dashed lines) and second order
(continuous line)  SWT
results. The Lanczos data are obtained on tilted lattices
$L \protect\sqrt{2} \times L \protect\sqrt{2}$
with $L=2$ (triangles), $L=3$ (squares) and $L=4$ (circles).
The lowest curves refers to the infinite size SWT  results.
\label{Fig. 1}}
\end{figure}

\begin{figure}
\caption{$a)$: rigid rotator anomaly
$\delta \chi\protect^{-1}=V (E(S)-E(0))$  \protect\linebreak
$-1/(2 \chi_L)$, where $\chi_L$ is estimated on finite sizes
 for 8 (triangles),
18 (squares) and 32 (circles) lattice sites. The continuous lines are
guides to the eye, the dashed line represents  the ideal  behaviour
in a quantum antiferromagnet.
$b)$: Finite size scaling of the inverse susceptibility. The finite
size data (notation as in Fig. 1)
extrapolated (see text) to infinite size (continuous line).
The stars are the exact values in the isotropic and one dimensional case.}
\label{Fig. 2}
\end{figure}

\begin{figure}
\caption{
Diagonalization data of the gap scaling function
(see Eq. \protect\ref{scala}) for the
two chain model, three chains and square clusters.
For the two and three chains,
open triangles refer to $L=4$,
full triangles $L=6$, open squares $L=8$, full squares $L=10$, open circles
$L=12$. For the square clusters open triangles correspond to 8 sites,
full triangles to 18 sites, open squares to 32 sites and stars to
the $4\times 4$ cluster.
\label{Fig. 3}}
\end{figure}
\begin{figure}
\caption{
 Square of the spin velocity ratio vs. anisotropy.
The dashed line is the leading SWT
result, the continuous line includes the one loop correction in the
thermodynamic limit.
Finite size estimates on a 32-site lattice for $\alpha Z(\alpha)$
are obtained by
exact diagonalization (full triangles) and second order finite size SWT
\protect\cite{zhong} (open triangles).}
\label{Fig. 4}
\end{figure}
\begin{figure}
\caption{
 Phase diagram of the spatially anisotropic Heisenberg model
obtained via one loop SWT. The order parameter vanishes along the
continuous line and the spin-wave velocity ratio along the dashed line.
The long dashed line indicates a crossover
transition between a decoupled phase (DP) and a normal disordered phase
with a finite spin velocity ratio.}
\label{Fig. 5}
\end{figure}


\begin{references}
\bibitem{faz} P. Fazekas, P.W. Anderson, Philos. Mag. {\bf 30},
423 (1974).
\bibitem{pwa} P.W. Anderson, Science, {\bf 235}, 1196, (1987).
\bibitem{fr} B. Bernu {\it et al.}  Phys. Rev. Lett. {\bf
69}, 2590 (1992).
\bibitem{schulz} H.J. Schulz, T. Ziman Europhys. Lett. {\bf 18},
355, (1992).
\bibitem{huse} P. W. Leung and V. Elser \prb {\bf 47} 5459 (1993).
\bibitem{oitmaa} J. Oitmaa, D.D. Betts, Can. J. Phys. {\bf 56}, 897
(1978).
\bibitem{ceperley} N. Trivedi, D.M. Cepeley, Phys. Rev. B {\bf 41},
4552 (1990).
\bibitem{runge} K.J. Runge, Phys. Rev. B {\bf 45}, 7229 (1992);
 {\it ibid},12292.
\bibitem{oguchi} T. Oguchi, Phys. Rev. {\bf 117}, 117 (1960).
\bibitem{bourb} See for instance C. Bourbonnais, L.G. Caron,
Int. J. Mod. Phys. {\bf 5}, 1033 (1991).
\bibitem{iome} A. Parola, Phys. Rev. B, {\bf 40}, 7109 (1989).
\bibitem{doucot} M. Azzouz, B. Doucot Phys. Rev. B, {\bf 47},
8660 (1992).
\bibitem{nelson} S. Chakravarty  {\it et al.}
Phys. Rev. Lett. {\bf 60}, 1057 (1988).
\bibitem{haldane} F.D.M. Haldane, J. Phys. C  {\bf 14}, 2585 (1981);
 Phys. Lett. {\bf 93A}, 464 (1983);  \prl {\bf 61}, 1029 (1988).
\bibitem{fisher} D.S. Fisher, Phys. Rev. B {\bf 39} 11783 (1989).
\bibitem{ziman} H. Neuberger, T. Ziman, Phys. Rev. B {\bf 39}, 2608 (1989).
\bibitem{zhong} Q.F. Zhong, S. Sorella, Europys. Lett. {\bf 21},
629 (1993).
\bibitem{strong} S.P. Strong, A.J. Millis, Phys. Rev. Lett.
{\bf 69}, 2419 (1992).
\bibitem{dag2c} T. Barnes {\it et al.} \prb {\bf 47} 3196 (1993).
\end{references}
\end{document}